# A New Model for Temperature Jump at a Fluid-Solid Interface

**Jian-Jun Shu\***, **Ji Bin Melvin Teo**, **Weng Kong Chan**

School of Mechanical & Aerospace Engineering, Nanyang Technological University, 50 Nanyang Avenue, Singapore 639798.

\* Corresponding author

E-mail: mjjshu@ntu.edu.sg (JJS)

**Competing Interests:** The authors have declared that no competing interests exist.

**Financial Disclosure:** The authors received no specific funding for this work.

## Abstract

The problem presented involves the development of a new analytical model for the general fluid-solid temperature jump. To the best of our knowledge, there are no analytical models that provide the accurate predictions of the temperature jump for both gas and liquid systems. In this paper, a unified model for the fluid-solid temperature jump has been developed based on our adsorption model of the interfacial interactions. Results obtained from this model are validated with available results from the literature.

## Introduction

Like the slip boundary condition [1], the temperature jump at a fluid-solid interface has a long history since its discovery by Smoluchowski [2]. The presence of an interfacial thermal resistance, known as the Kapitza resistance, was experimentally detected for liquid helium in a superfluid phase [3]. Little attention has been placed on the phenomena for liquid-solid interfaces until recently when the accessibility of micro and nanoscale fabrication and molecular simulations motivated researchers to explore the feasibility of temperature jump occurring under room conditions.

For gas-solid interfaces, the existing temperature jump models largely follow from the kinetic theory derivation of the slip velocity with the use of the thermal accommodation coefficient, $\sigma_T$, which represents the fraction of reflected or re-emitted molecules possessing the mean energy of gas molecules at the same temperature as the wall [4]. A thermal accommodation coefficient of one may be interpreted as molecule undergoing repeated collisions with the wall and finally getting re-emitted as if it were from a gas at the wall temperature. In contrast, a molecule that is reflected immediately on impact can be thought of as having a thermal accommodation coefficient of zero. In effect, the thermal accommodation coefficient merely categorises molecules into those that fully equilibrate to the energy of the wall and those that retain their original energy. Later models have considered other factors such as intermolecular interactions, molecular velocity distribution and angles of incidence of the impinging gas molecules [5-7]. The thermal accommodation coefficient is assumed to be a constant in most studies but experimental measurements have reflected a dependence on the wall temperature. The existing theoretical model of the liquid-solid temperature jump adopts the continuum phonon-scattering formulation but is only valid at extremely low temperatures and neglects the influence of molecular interactions at the boundary [8,9]. To the best of our knowledge, there are no analytical models that provide the accurate predictions of the temperature jump for both gas and liquid systems. In this paper, a unified model for the fluid-solid temperature jump has been developed based on our adsorption model of the interfacial interactions.

# Materials and Methods

## Interfacial temperature jump from fluid-solid molecular interactions

We consider a quiescent fluid layer that resides on a solid surface in the presence of an externally applied temperature gradient or heat source. In the absence of corrugations, this restricts the interactions to the components of kinetic energy normal to the surface since the parallel components effectively cancel out for an equilibrium distribution.

### Mean kinetic energy of surface fluid particles

Without driven flow, the particles in the mobile and inelastically desorbed states can be jointly grouped into the precursor state where the probability of a particle being in this state is $p_p$. In the absence of an external force, the surface hopping of the mobile particles has the characteristics of a symmetrical random walk with a zero mean drift. A schematic illustration of the adsorption and desorption states is shown in Fig 1.

**Fig 1.** Energies of particles in the following states: (a) incident (b) elastic scattering (c) pre-cursor (d) desorbed

The mean kinetic energy of the surface fluid particles $E_s$ can be expressed as

$$E_s = (1-p_s)E_e + p_s p_p E_p + p_s(1-p_p)E_{des} \qquad (1)$$

where $E_e$, $E_p$ and $E_{des}$ refer to the respective kinetic energy of particles that are elastically scattered, in the precursor and desorbed states and $p_s$ refers to the sticking probability [10].

Elastically scattered particles retain their incident kinetic energy prior to impact $E_i$

$$E_e = E_i. \qquad (2)$$

Particles that are trapped in the precursor state experience a loss in energy upon impact that is sufficiently large to prevent them from escaping back to the bulk fluid immediately while still preventing them from falling into the bottom of the potential well. We introduce the coefficient of restitution $\varepsilon$ that represents the ratio of pre- and post-impact thermal velocities. Hence the kinetic energy of a particle in the precursor state is given by

$$E_p = \frac{\varepsilon^2}{2} E_i. \qquad (3)$$

The desorbed particles, having spent a residence time longer than that required for equilibration, attain the thermal equilibrium with the surface and therefore emerge with kinetic energy that is the characteristic of particles possessing the temperature of the surface

$$E_{des} = E_w \qquad (4)$$

where $E_w$ denotes the kinetic energy of particles at the temperature of the solid surface. Putting all the energy terms together, we can replace the kinetic energy terms by the temperatures as well as temperature gradients to derive the final functional form of the temperature jump expression.

## General temperature jump boundary condition

The substitution of Eqs 2 to 4 into Eq 1 and rearranging allows us to obtain the following form for the energy balance

$$E_s - E_i = p_s\left[\left(1 - p_p \frac{\varepsilon^2}{2}\right)(E_w - E_i) - p_p\left(1 - \frac{\varepsilon^2}{2}\right)E_w\right]. \qquad (5)$$

The difference in kinetic energy between the incident and surface particles on the left-hand side of Eq 5 can be expressed in terms of the thermal energy conducted between the fluid and solid

$$E_s - E_i = kA\tau \left.\frac{dT}{dn}\right|_s \qquad (6)$$

where $\left.\frac{dT}{dn}\right|_s$ refers to the fluid temperature gradient at the surface, $k$ is the thermal conductivity of the fluid, $A$ is the effective surface area of thermal conduction and $\tau$ is the characteristic sticking time.

The kinetic energy difference in the first-term on the right can be approximated [11] by

$$E_w - E_i \approx \alpha k_B\left(T_w - T_s + C_\lambda \left.\frac{dT}{dn}\right|_s\right) \qquad (7)$$

where $T_w$ and $T_s$ refer to the fluid temperatures at the wall and that at one temperature jump distance $C_\lambda$ away, $k_B$ is the Boltzmann constant and $\alpha$ is a factor that accounts for the number of molecular degrees of freedom being considered *i.e.* translational, rotational or vibrational. For instance, $\alpha$ takes on a value of 2 for pure translational motion if the rotational and vibrational degrees of freedom are neglected [4].

The substitution of the kinetic energy terms in Eqs 6 and 7 into Eq 5 gives the final form of the temperature jump as

$$T_s - T_w = C_1 \left.\frac{dT}{dn}\right|_s - C_2 T_w \qquad (8)$$

where the coefficients $C_1 = C_\lambda - \frac{2kA\tau}{\alpha k_B p_s (2 - p_p \varepsilon^2)}$ and $0 \leq C_2 = \frac{p_p (2 - \varepsilon^2)}{2 - p_p \varepsilon^2} \leq 1$ represent the interfacial conditions, adsorption probabilities and properties of the media.

The temperature jump expression in Eq 8 marks a new model for the temperature discontinuity at a fluid-solid interface that has been derived based on adsorption theory. Though the general trend of the temperature jump behaviour with respect to the temperature gradient remains largely similar, the temperature jump coefficient $C_1$ differs slightly from the original model for gas-solid interfaces by Smoluchowski (and other adaptations) due to the introduction of a trapping phase. Molecular interactions are also explicitly considered in the new model, unlike the acoustically based Kapitza resistance models for liquid-solid interfaces. The inclusion of a precursor state also produces an additional dependence on the surface temperature $T_w$, the second term on the right hand side of Eq 8, which may explain the experimental observations of the surface temperature dependence of the thermal accommodation coefficient as well as thermal rectification effects found in the simulations of heat transfer of liquid-solid systems that have been reported in the literature.

# Results

## Validation of new temperature jump boundary condition

A review of the literature shows that few experimental temperature jump studies have been carried out in recent years. The main difficulty lies in the measurement of temperatures of the solid and fluid at the interface within enclosed setups, which researchers have attempted to circumvent using indirect measurement techniques. Using modern apparatus, researchers have revisited traditional temperature jump experimental setup for gases to acquire the higher-resolution measurement of the thermal accommodation coefficient. In the study of liquid-solid thermal boundary resistance, molecular dynamics (MD) simulation is the preferred tool of choice with only one experimental measurement of room-temperature liquid being reported till date.

### Experimental measurement of gas-solid temperature jump

For gas-solid interfaces, two experimental studies have been selected based on the findings of wall temperature dependent temperature jump coefficients which cannot be predicted using the conventional temperature jump model due to the assumption of a constant thermal accommodation coefficient.

Hall & Martin [12] obtained the value of thermal accommodation coefficient from the measurement of the thermal conductivity of $UO_2$ beds that were packed between two concentric cylinders and filled with the test gases. Yamaguchi *et al*. [13]

measured the heat flux in a refined setup of the traditional coaxial cylinder system under rarefied conditions to perform the updated measurement of thermal accommodation coefficient.

## Comparison of new model with experimental data for gas-solid interface

Theoretically, the thermal accommodation coefficient in the free-molecular regime is derived from the expression

$$\sigma_T = \frac{E_s - E_i}{E_w - E_i}. \qquad (9)$$

The substitution of Eqs 6 to 8 into the above equation allows us to derive the following form of a temperature-dependent thermal accommodation

$$\sigma_T = \frac{1}{a + bT_w} \qquad (10)$$

where $a = \dfrac{2}{p_s(2 - p_p \varepsilon^2)}$ and $b = \dfrac{\alpha k_B p_p (2 - \varepsilon^2)}{kA\tau(2 - p_p \varepsilon^2)\left.\dfrac{dT}{dn}\right|_s}$. The theoretical predictions using Eq 10 of the experimentally measured thermal accommodation coefficients from Hall & Martin [12] and Yamaguchi *et al.* [13] are shown in Figs 2-4.

**Fig 2.** Temperature dependence of thermal accommodation coefficient of UO$_2$ sphere beds in helium. Symbols: Experimental data [12]. Line: Theoretical prediction using Eq 10 with $a = 1.582$ and $b = 1.319 \times 10^{-3}$.

**Fig 3.** Temperature dependence of thermal accommodation coefficient of UO$_2$ sphere beds in argon. Symbols: Experimental data [12]. Line: Theoretical prediction using Eq 10 with $a = 0.828$ and $b = 1.702 \times 10^{-4}$.

**Fig 4.** Temperature dependence of the thermal accommodation coefficient for a platinum-argon interface for $10 < k_n < 250$. Symbols: Experimental data [13]. Line: Theoretical prediction using Eq 10 with $a = 0.604$ and $b = 1.447 \times 10^{-3}$.

## Measurement of liquid-solid temperature jump

Temperature jump for liquid-solid interface has been measured using a time-domain thermoreflectance technique [14]. Unfortunately, the lack of temperature jump data in the published report did not allow any meaningful comparisons. Here, four MD simulation studies conducted by separate groups are used for the corroboration of our new temperature jump model.

Kim *et al.* [15] performed the MD simulations of steady state heat conduction between the parallel plates with nanoscale gaps filled with liquid argon. Shenogina *et al.* [16] studied the effect of wetting on thermal conductance for the different interfaces of self-assembled monolayer (SAM) and water. Hu *et al.* [17] conducted the nonequilibrium MD heat conduction simulations of a system consisting of SAM bonded to a silica surface that was submerged within a water phase. The nonequilibrium MD simulations of Acharya *et al*. [18] involved the study of the Kapitza thermal conductance of solid-liquid interfaces between SAM and liquid water for mixed -$CF_3$/-OH SAMs.

## Comparison of new model with MD simulation data for liquid-solid interfaces

The temperature jumps versus wall temperature gradient curves from the four sets of MD simulations are replicated in Figs 5-8. The wall temperature gradients were evaluated using the Fourier's heat conduction law for given heat fluxes. On the same graphs, the theoretical prediction using Eq 8 of the experimentally measured temperature jump is plotted. Also shown in the figures are predictions obtained using the existing temperature jump model

$$T_s - T_w = C \frac{dT}{dn}\bigg|_s \qquad (11)$$

where $C$ represents the temperature jump coefficient, also referred to as the Kapitza length in the literature.

**Fig 5.** Temperature jump as a function of wall temperature gradient at a solid-liquid argon interface. Symbols: MD simulation results at $T_w = 160K$ (triangles), $T_w = 90K$ (circles) [15]. Solid line: New temperature jump model from Eq 8 with $C_1 = 2.348 \times 10^{-9}$ and $C_2 = 0.036$ for $T_w = 160K$, $C_1 = 2.121 \times 10^{-9}$ and $C_2 = 0.081$ for $T_w = 90K$. Dashed line: Existing temperature jump model from Eq 11 with $C = 1.623 \times 10^{-9}$ for $T_w = 160K$, $C = 1.207 \times 10^{-9}$ for $T_w = 90K$.

**Fig 6.** Temperature jump as a function of wall temperature gradient at a SAM-water interface. Symbols: MD simulation results for hydrophobic -$CF_3$ SAM (triangles) and hydrophilic -OH SAM (circles) [16]. Solid line: New temperature jump model from Eq 8 with $C_1 = 6.121 \times 10^{-9}$ for -$CF_3$ SAM at $T_w = 300K$, $C_1 = 1.525 \times 10^{-9}$ for -OH SAM at $T_w = 285K$.

**Fig 7.** Temperature jump as a function of wall temperature gradient at a silica-SAM-water interface. Symbols: MD simulation results [17]. Solid line: New temperature jump model from Eq 8 with $C_1 = 1.04 \times 10^{-9}$ and $C_2 = 9.682 \times 10^{-4}$ for $T_w = 292K$. Dashed line: Existing temperature jump model from Eq 11 with $C = 1.007 \times 10^{-9}$.

**Fig 8.** Temperature jump as a function of wall temperature gradient at a silica-SAM-water interface. Symbols: MD simulation results [18]. Solid line: New temperature jump model from Eq 8 with $C_1 = 3.59 \times 10^{-9}$ and $C_2 = 0.015$ for $T_w = 326K$. Dashed line: Existing temperature jump model from Eq 11 with $C = 2.704 \times 10^{-9}$.

# Discussion

As seen in Figs 2-4, the prediction by the new temperature jump model displays good agreement with the results of the two reference experiments for gas-solid interfaces. This is due to the fact that the new model is able to reflect the wall temperature dependent behaviour of the thermal interactions that lead to the temperature discontinuity at the interface whereas the conventional temperature jump models assume a constant thermal accommodation coefficient.

Interestingly, in Fig 3, the measured thermal accommodation coefficient for argon gas is above unity. Hall & Martin [12] postulated from a kinetic theory perspective that this over-accommodation could be attributed to surface roughness which promotes more efficient heat exchange between the gas molecules and solid surface due to the higher tendency for the gas molecules to be scattered at larger angles and therefore remained within the vicinity of the surface. Based on the definition of the thermal accommodation coefficient, it connotes that the net energy exchange is greater than the available difference in energy, which appears to violate the second law of thermodynamics. Furthermore, argon is a monatomic gas and therefore should not experience an exchange of energy modes with the internal degrees of freedom. This leads to the point that the thermal accommodation coefficient by itself may not provide an adequate description of the molecular interactions at the surface using a straightforward specular and diffuse reflection model. Hence, it should not be inferred from the variation of the thermal accommodation coefficient with temperature that the thermal accommodation coefficient is a function of the temperature. Rather, it could be explained by more complex forms of molecule-surface interactions, such as the precursor adsorption states considered in our model which the gas molecules may assume upon impacting the surface, consequently contributing to the temperature discontinuity at the interface.

From the comparisons of the agreement between the analytical curves and experimental data for liquid-solid interfaces displayed in Figs 5-8, the new temperature jump model described by Eq 8 ostensibly offers a better prediction over that of the existing model. In particular, it can be observed that the temperature jump in most of the MD simulation results does not vanish when the wall temperature gradient decreases to zero. This suggests that the temperature jump is not merely driven by the fluid temperature gradient but also affected by the thermal energy of the solid molecules.

The experimental data of Shenogina *et al*. [16] in Fig 6 depicts the contrasting temperature jump behaviours of hydrophilic and hydrophobic surfaces. Granted that the conventional temperature jump model is able to provide a good prediction of the experimental data using different temperature jump coefficients, we can provide a qualitative explanation of the influence of wetting using our new model since the sticking time is expected to decrease with increasing hydrophobicity. Indeed, for the hydrophobic $CF_3$ SAM, the value of $C_1$ is $6.121 \times 10^{-9}$ while that for the hydrophilic OH SAM is $1.525 \times 10^{-9}$, corresponding to a higher sticking time with reference to Eq 8. This is also supported by the lower value of $C_1 = 3.59 \times 10^{-9}$ for the hydrophilic–$CONH_2$ surface studied by Acharya *et al*. [18] in Fig 8.

At elevated temperature gradients, the experimental data begins to deviate from linear behaviour predicted by both the conventional and new temperature jump models, instead displaying a non-linearly decreasing tail that draws parallels with the shear rate dependence of the slip length at increased wall shear rates. It is noted that only one group reported similar non-linear findings from their MD simulations of a silicon-water system [19]. However, in their case, the temperature jump increased non-linearly with increasing heat flux. Owing to the paucity of available data, the non-linear behaviour warrants further investigation.

The MD simulation of heat transfer across liquid-solid interfaces have unveiled a thermal rectification effect, whereby the magnitude of the temperature jump changes with the direction of the heat flux for the same absolute value. Our new model reflects this phenomenon which has several potential uses such as thermal diodes or temperature cloaks. The thermal rectification effect is graphically depicted in Fig 9 using similar values for the wall temperature in Eq 8. It can be observed that a heat current flowing from the liquid phase to solid phase diminishes the magnitude of the temperature jump while reversing the direction results in an augmented temperature jump. The closer inspection of the temperature distributions in certain MD simulations purporting this rectification property reveals a difference in wall temperatures when the direction of heat flux is altered. For example, the wall temperatures differ by $23K$ and $16K$ respectively in the simulations of Hu *et al*. [17] and Acharya *et al*. [18]. However, we note that the temperature jump in the case of the former increases at a steeper rate when the direction of heat flux points from the solid to liquid. According to our model, this wall temperature disparity may possibly give rise to an apparent rectification effect since the magnitude of the temperature jump is affected by the boundary temperature. The stricter control of the interfacial temperature is necessary in order to rule out its influence on the resultant temperature jump. The temperature jump data shown in Fig 5 provides the evidence of this wall temperature dependence for two wall temperatures of $160K$ and $90K$, though Kim *et al*. [15] did not claim to have observed the rectification effect.

**Fig 9.** Thermal rectification effect with a change in direction of heat flux. A negative temperature gradient refers to decreasing fluid temperatures with increasing normal distance from the solid surface and vice versa for a positive temperature gradient.

## Conclusions

In summary, we have developed a general model that is capable of describing the temperature discontinuity across a fluid-solid interface based on the energy balance of fluid molecules in various adsorption states. The applicability of the model to both gas and liquid systems is substantiated by the good agreement with experimental data from the literature. In particular, the wall temperature dependence of the thermal accommodation coefficient, which is assumed to be constant in majority of the gaseous temperature jump studies, is well-represented by the model. The improved predictions of experimental measurements of liquid-solid temperature jump are also obtained using the new model.

## Author Contributions

**Conceptualization:** JJS JBMT.

**Methodology:** JJS JBMT.

**Software:** JBMT.

**Validation:** JJS JBMT.

**Formal analysis:** JJS JBMT.

**Investigation:** JJS JBMT.

**Resources:** JJS JBMT WKC.

**Data curation:** JBMT.

**Writing (original draft preparation):** JJS JBMT.

**Writing (review and editing):** JJS JBMT.

**Visualization:** JBMT.

**Supervision:** JJS WKC.

**Project administration:** JJS WKC.

**Funding acquisition:** JJS WKC.

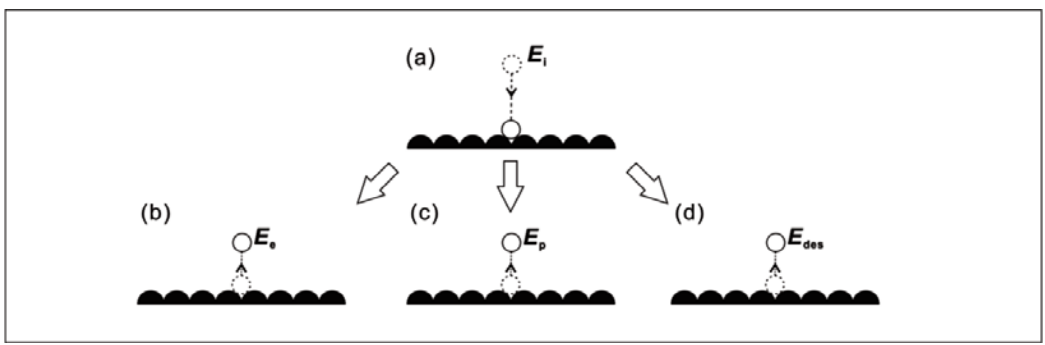

**Fig 1.** Energies of particles in the following states: (a) incident (b) elastic scattering (c) pre-cursor (d) desorped.

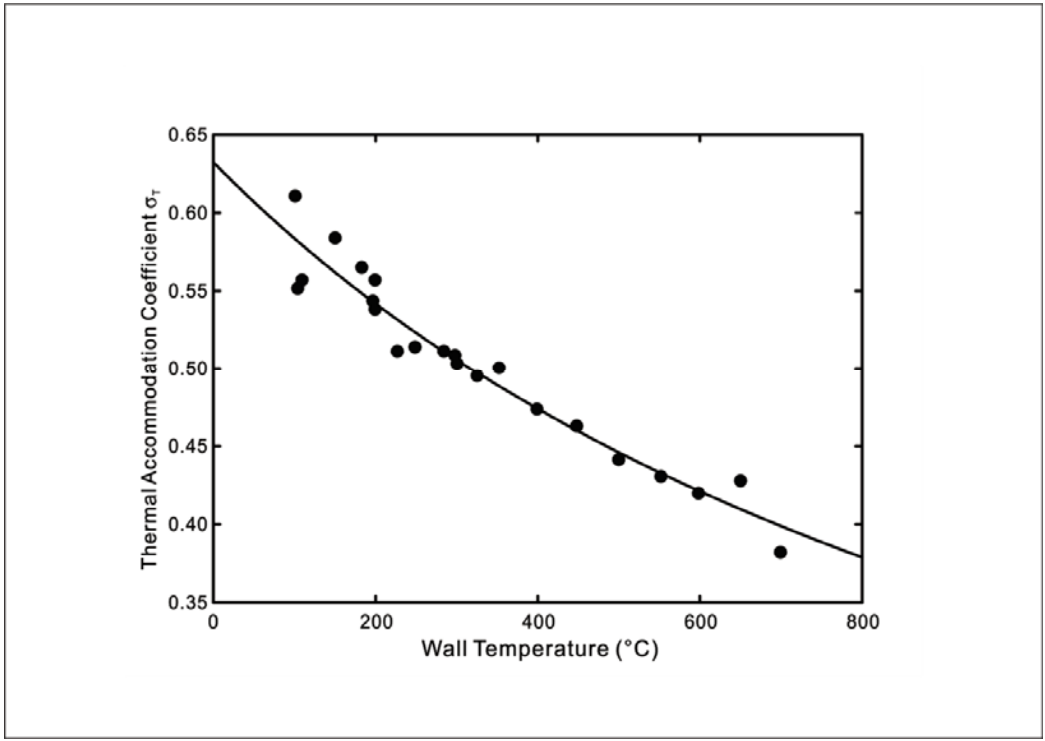

**Fig 2. Temperature dependence of thermal accommodation coefficient of UO$_2$ sphere beds in helium.** Symbols: Experimental data [12]. Line: Theoretical prediction using Eq 10 with $a$ = 1.582 and $b$ = 1.319 × 10$^{-3}$.

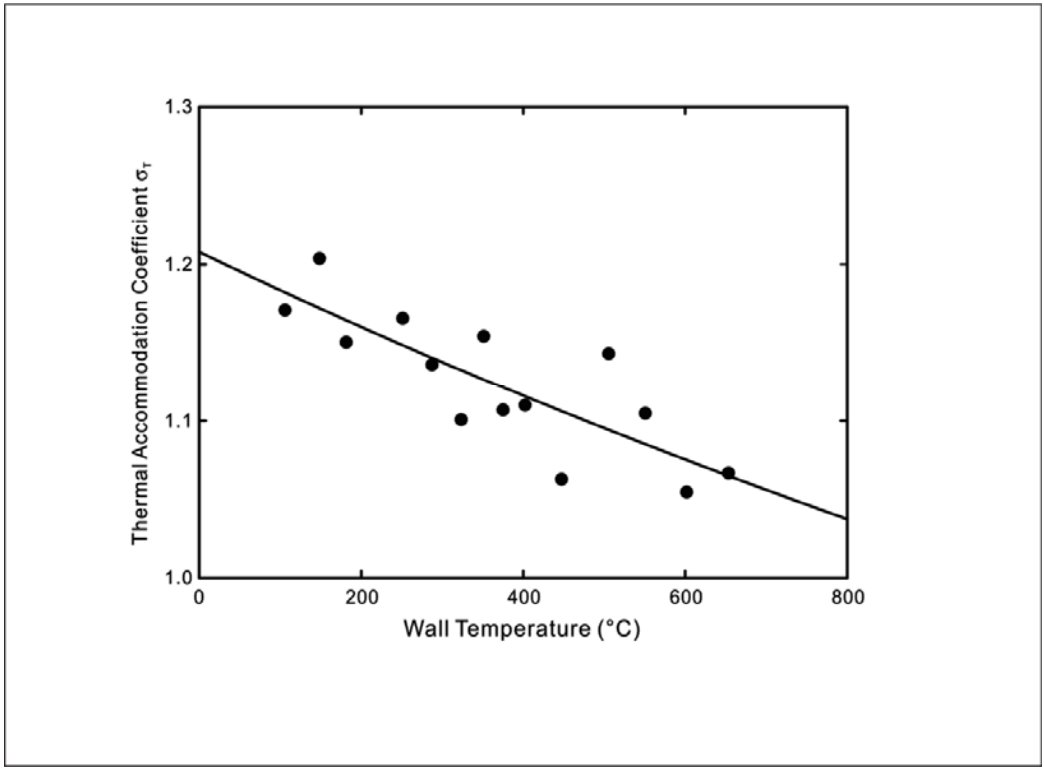

**Fig 3. Temperature dependence of thermal accommodation coefficient of UO$_2$ sphere beds in argon.** Symbols: Experimental data [12]. Line: Theoretical prediction using Eq 10 with *a* = 0.828 and *b* = 1.702 × 10$^{-4}$.

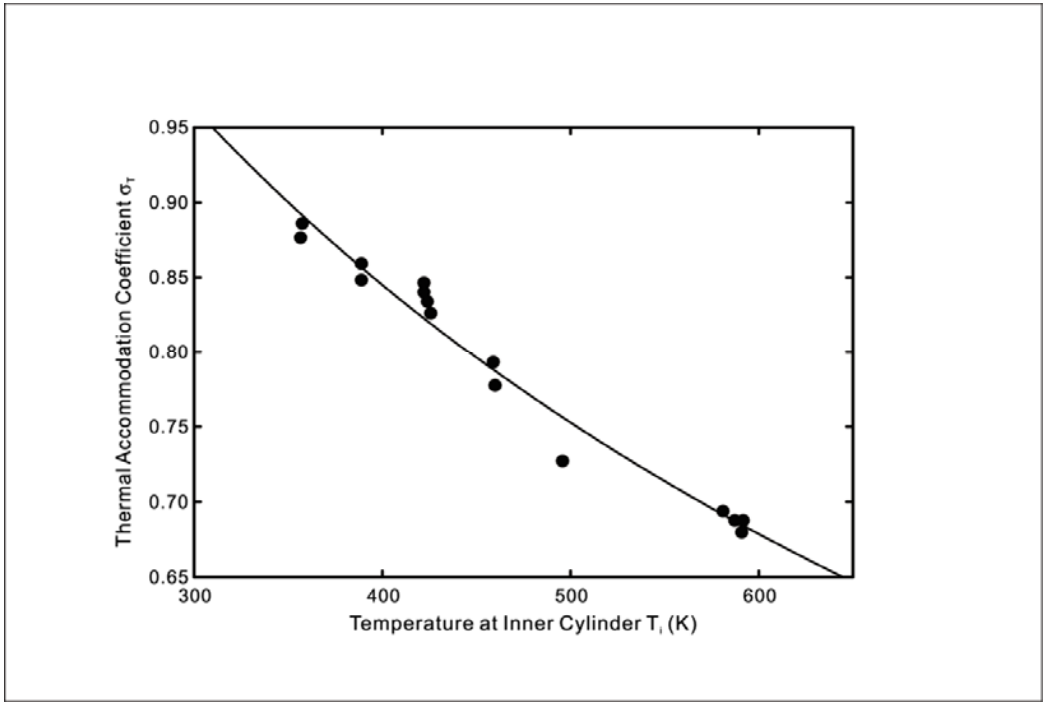

**Fig 4. Temperature dependence of the thermal accommodation coefficient for a platinum-argon interface for $10 < k_n < 250$.** Symbols: Experimental data [13]. Line: Theoretical prediction using Eq 10 with $a = 0.604$ and $b = 1.447 \times 10^{-3}$.

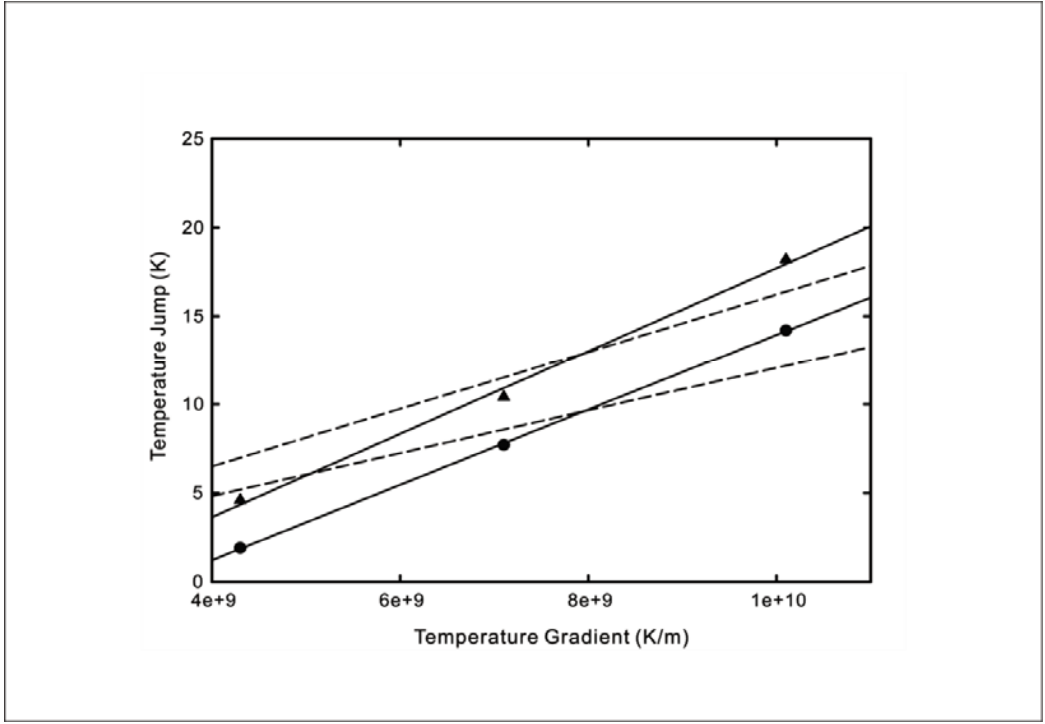

**Fig 5. Temperature jump as a function of wall temperature gradient at a solid-liquid argon interface.** Symbols: MD simulation results at $T_w$ = 160K (triangles), $T_w$ = 90K (circles) [15]. Solid line: New temperature jump model from Eq 8 with $C_1$ = 2.348 × $10^{-9}$ and $C_2$ = 0.036 for $T_w$ = 160K, $C_1$ = 2.121 × $10^{-9}$ and $C_2$ = 0.081 for $T_w$ = 90K. Dashed line: Existing temperature jump model from Eq 11 with $C$ = 1.623 × $10^{-9}$ for $T_w$ = 160K, $C$ = 1.207 × $10^{-9}$ for $T_w$ = 90K.

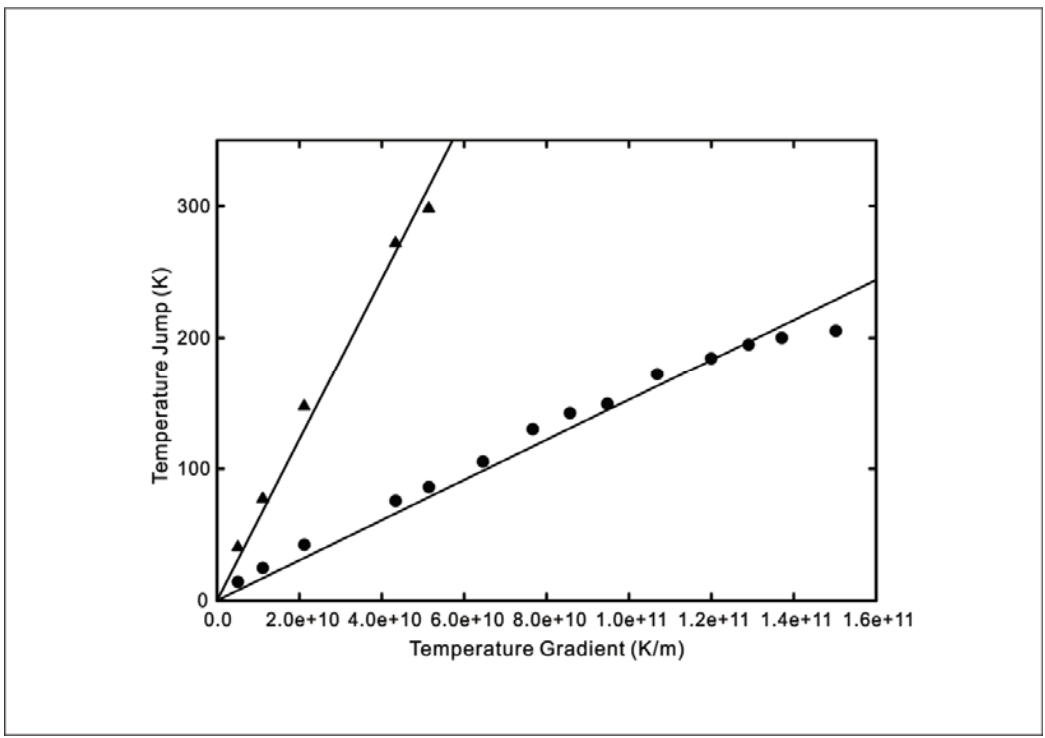

**Fig 6. Temperature jump as a function of wall temperature gradient at a SAM-water interface.** Symbols: MD simulation results for hydrophobic -CF$_3$ SAM (triangles) and hydrophilic -OH SAM (circles) [16]. Solid line: New temperature jump model from Eq 8 with $C_1 = 6.121 \times 10^{-9}$ for -CF$_3$ SAM at $T_w = 300K$, $C_1 = 1.525 \times 10^{-9}$ for -OH SAM at $T_w = 285K$.

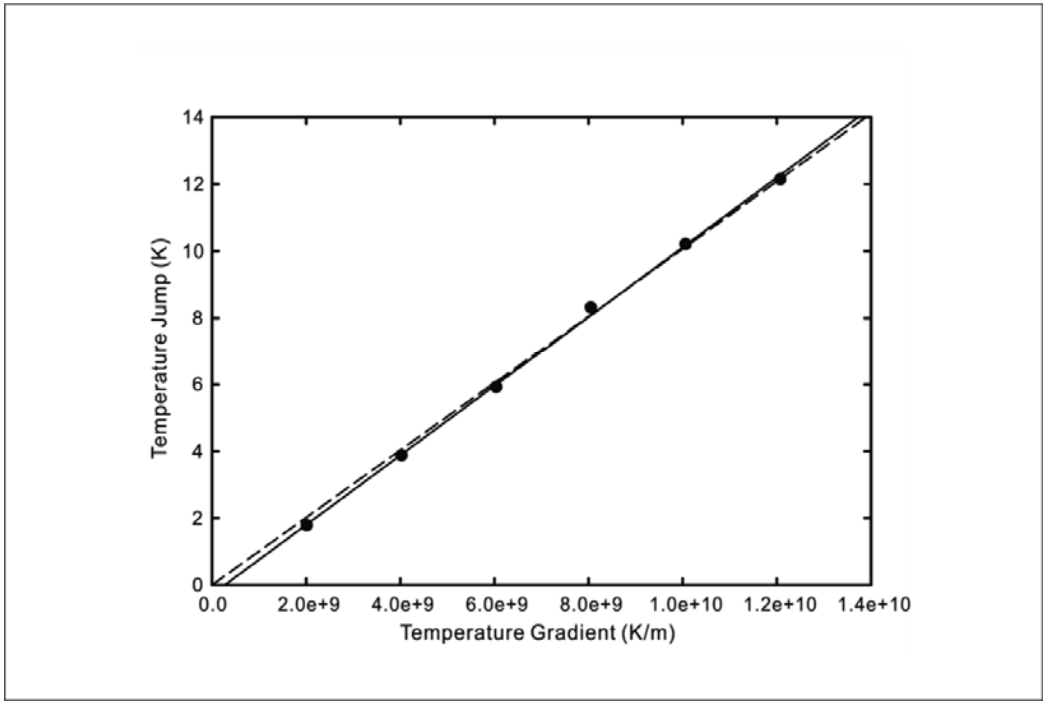

**Fig 7. Temperature jump as a function of wall temperature gradient at a silica-SAM-water interface.** Symbols: MD simulation results [17]. Solid line: New temperature jump model from Eq 8 with $C_1 = 1.04 \times 10^{-9}$ and $C_2 = 9.682 \times 10^{-4}$ for $T_w = 292K$. Dashed line: Existing temperature jump model from Eq 11 with $C = 1.007 \times 10^{-9}$.

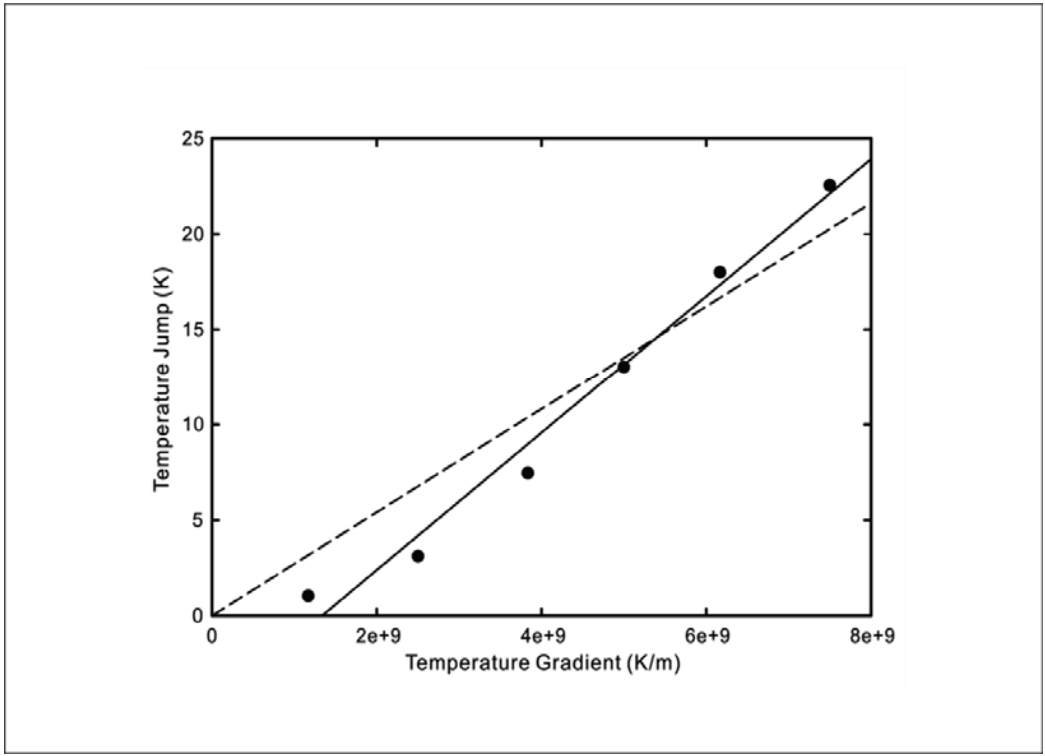

**Fig 8. Temperature jump as a function of wall temperature gradient at a silica-SAM-water interface.** Symbols: MD simulation results [18]. Solid line: New temperature jump model from Eq 8 with $C_1 = 3.59 \times 10^{-9}$ and $C_2 = 0.015$ for $T_w = 326K$. Dashed line: Existing temperature jump model from Eq 11 with $C = 2.704 \times 10^{-9}$.

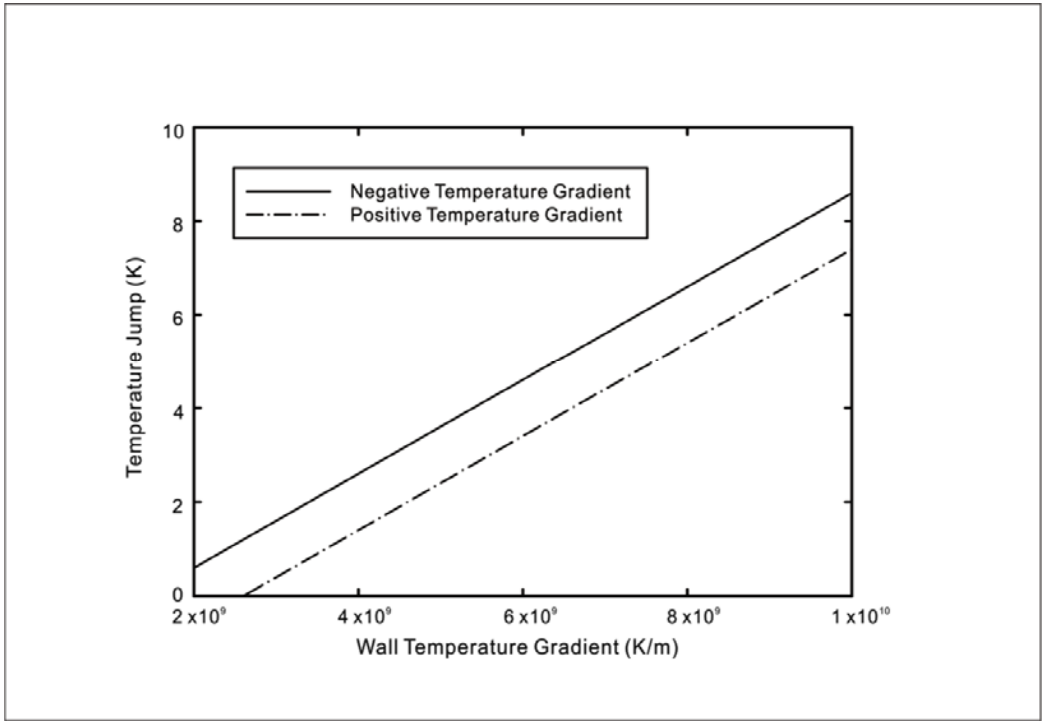

**Fig 9. Thermal rectification effect with a change in direction of heat flux.** A negative temperature gradient refers to decreasing fluid temperatures with increasing normal distance from the solid surface and vice versa for a positive temperature gradient.